\documentstyle[twoside,j1tp]{article}
\title{ On the Classification of the Low Energy Excitations
in the Doped  Antiferromagnet}
\author{Sandro Sorella \address{International School for Advanced Studies \\
Via Beirut 2-4 I-34013 Trieste Italy}}
\runninghead{Sandro Sorella}{On the Classification of the Low Energy
 Excitations in an Antiferromagnet}
\begin{document}
\begin{abstract}
 A  long-wavelength, low energy hamiltonian is derived to
describe the dynamic of a single hole in a quantum antiferromagnet in
two or higher  spatial dimensions.
 In this {\em exactly solvable} limit a {\em new} kind of symmetry
 is important
to classify the elementary spin excitations of a single hole in the $t-J$
model.
This symmetry is hidden  at finite size or at short wavelength.
The resulting classification has important consequences for understanding
the physics of doped antiferromagnets.
 Fermi liquid like $s=1/2$ and
charge one excitations are still well defined in a quantum antiferromagnet,
but are not a {\em complete} set to describe the low energy physics.

\end{abstract}
\maketitle

\section{INTRODUCTION}
Within Fermi liquid theory  the elementary low energy
excitations of an interacting gas of electrons are classified, by assuming
that   the many -body quantum states of the interacting  system
are in correspondence with  the free electron states of a
Fermi gas, by -say- an adiabatic increase  of the interaction.
Following this assumption  many of the thermodynamic and
dynamical properties  of the quantum liquid are consistently derived, from the
finite spin  susceptibility at zero temperature to the zero sound propagation.

In this contribution  I will derive the large spin  (low energy), long
wavelength  hamiltonian for a single charge excitation in an antiferromagnet
(AF),
 but allowing for arbitrary spin excitations.
   I will show that in a doped AF
 still Fermi liquid excitations that carry charge one and spin
$1/2$ are well defined elementary excitations in the thermodynamic limit.
However  beyond  these conventional excitations also collective low energy
excitations  exist that change the spin of the many body system without
changing
the charge. They are also {\em elementary} excitations (and they may be called
spin-wave spinons\cite{andersonr})  since they are in correspondence with
the spin one elementary excitations of the 2D quantum antiferromagnet, namely
the spin-waves.
Thus the Fermi liquid set of states turns to be an incomplete classification
in a doped antiferromagnet and this may  represent the key-feature to explain
the  anomalous properties of the high $T_c$ superconductors.

I consider  a single hole in
 the $t$-$J$ Hamiltonian  for spatial dimension $d \ge 2$:
$$
H_{tJ} =  t \sum \limits_{<  i,j > ,\sigma } {\cal P}
(c^{\dag}_{i,\sigma} c_{j,\sigma} + h.c. ) {\cal P}  +
 J  \sum \limits_{<  i,j >} (\vec S_i \, \vec S_j - {1\over 4 } n_i \,
n_j), $$
where $< i,j >$ denotes a summation over the nearest neighbor
in a finite lattice with $L$ sites and periodic boundary conditions,
 ${\cal P}$ is the projector onto the Hilbert space without
doubly occupied sites, $\vec S_i $ and $n_i$ are the electron spin and number
operators at site $i$, and $c^{\dag}_{i,\sigma}$ and $c_{i,\sigma}$ are the
usual creation and  annihilation operators for electrons of spin $\sigma$.
The results presented here should not depend on the details of the model
since at long-wavelength and low energy universal behavior is expected.

\section{Galileo transformation for the hole states.}\cite{xiang}  In the
reference  frame where the hole is at the origin $O$ of the coordinates, the
most general one hole state  with momentum $p$ reads:
\begin{equation}
|\psi_p> \,=\,{1\over \sqrt L } \sum \limits_{R,\sigma}
e^{i p R} c_{R,\sigma} T_R  |S_O>
\label{psieff}
\end{equation}
where $|S_O>$ is a pure spin state that satisfies:
i) $n_R |S_O> = |S_O>$ for all sites $R$,  ii)
$c^{\dag}_{O,\downarrow} c_{O,\downarrow}  |S_O> \,=\,  |S_O>$, i.e. the spin
at
the origin is  fixed  to $-1 \over 2$.
Here $T_{R}$ is the translation operator  by a vector $R$, i.e.
$T_R \vec S_{R^{\prime}}  T_{-R} = \vec S_{R+R^\prime}$.

It is simple to show that any one hole wavefunction, expressed
in the form (2), is an eigenstate of $H_{t-J}$ with momentum $p$,if the spin
wavefunction $|S_O>$  is  an eigenstate of the following spin
hamiltonian:\cite{zhongnew}
\begin{equation}
 H^{eff}_p \,=\, J d +  \sum \limits_{\tau_{\mu}}
\chi_{O,\tau_{\mu}} ( t e^{i p \tau_{\mu}} T_{\tau_{\mu}} -{J \over 2})
\,+\, H_{SW}
\label{heff}
\end{equation}
where $\chi_{R_i,R_j}= 2 \vec S_{R_i} \cdot \vec S_{R_j} + {1 \over 2}$
permutes two spins at sites $R_i$ and $R_j$, $\tau_{\mu}$ are nearest neighbor
 displacements  and $H_{SW}= J \sum \limits
_{<i,j>} (\chi_{i,j}-1)/2 $ is the well known translation invariant
Heisenberg model.
As expected after the Galileo transformation one gets a modified Heisenberg
hamiltonian in presence of a localized
perturbation.\cite{xiang,sorellaold}
 For example for $t=0$  the bonds
around the hole cannot contribute  to the magnetic energy and $H^{eff}_p$
differs from $H_{SW}$ only around the origin.

  This formalism
is very powerful because each eigenstate of $H^{eff}_p$ can be seen as an
adiabatic modification of an eigenstate of the Heisenberg hamiltonian $H_{SW}$
which I assume in the following to be well described by spin-wave
theory for $d\ge 2$.\cite{anderson,nelson,sorella}

 The effective spin hamiltonian (\ref{heff}) commutes with the spin at the
origin  $\vec S_O$ and the total spin $\vec S$. Since  the latter quantity
is defined also for $H_{SW}$, it is  convenient to
use the total spin $S$ and its $z-$ component $S_z$ as good quantum numbers for
the  eigenstates $|S, S_z>$ of (\ref{heff}).
 In order to define a  one hole eigenstate in (\ref{psieff}) it is enough to
 project each eigenstate $|S,S_z>$  onto  the ones with
definite $S^z_O$, i.e $|S_O>=({1\over2} -S^z_O ) | S, S^z >$.   Note
that, after this projection, the spin of the hole eigenstate  (\ref{psieff}) is
  $|S\pm 1/2|$.
Thus the correspondence of the one hole states
with the ones  of the spin
hamiltonian (\ref{heff}) with total spin $S$ is unique only in  the singlet
subspace where  I restrict the following analysis.
 In this  subspace- obviously important for an AF- we can use
  the  total spin $S$ to classify the spin of the
elementary excitations, analogously to what was done in the 1D Heisenberg model
where the spinons have been found to carry spin ${1\over 2}$.\cite{faddev}

The hamiltonian (\ref{heff}) is exact , and  the presence of the translation
operator makes  difficult to use standard approaches  as for the simpler
$H_{SW}$. The recent ansatz\cite{siggia}  proposed by  Shraiman and Siggia
corresponds to a  variational semiclassical solution  of the hamiltonian
(\ref{heff}),
 yielding for
example  the N\'eel state for $H_{SW}$.
However in the one hole case the semiclassical solution cannot be controlled
by the small parameter $1/s$, important to derive the spin-wave limit
for the Heisenberg model.\cite{anderson}

\section{ $J=2t$ Exact eigenstates for any $d$}.
 In this case  the kinetic part proportional to $t$
is exactly canceled by the magnetic bonds around the origin in (\ref{heff}). In
fact any eigenstate $|S,S_z>_H$ of the Heisenberg hamiltonian $H_{SW}$
with total ($spinon$) momentum $Q=p$ (i.e. $T_{\tau_\mu} |S,S_z>_H=e^{-i p
\tau_{\mu}} |S,S_z>_H $ ) is an exact eigenstate of (\ref{heff}) with the same
$p$. Although the  states with all possible  spinon momentum $Q=p$ are not a
complete set for the one hole Hilbert space,  it is remarkable  that for $J=2t$
a
considerable  fraction ($\sim {1\over L} $) of all the eigenstates is exactly
known and indeed coincide with the ones of $H_{SW}$. There is also numerical
evidence that such fraction of states contains the ground
state for vanishing momentum.\cite{antimo}

 The strict relation between the one-hole eigenstates with the ones of the
Heisenberg model  gives a robust evidence that the elementary excitations of
the
doped  antiferromagnet have to be classified in terms of $H_{SW}$ and not with
the free electron gas model as in the Fermi liquid theory. Fermi liquid
classification would hold only if the elementary excitations have spin $1/2$
and
charge one, implying  that in the effective hamiltonian (\ref{heff}) only the
ground state with $S=0$ should  be important in the low energy limit.  Instead
the hamiltonian (\ref{heff}) almost certainly has not a gap to the first
excited state because , the occurrence of long range  magnetic order -now
commonly accepted for $H_{SW}$ in $d=2$ and {\em rigorous} in $d=3$\cite{lieb}-
implies a broken $SU(2)$ symmetry and the existence of gapless Goldstone modes.

\section{ $t=0$ Static limit } Here
the hamiltonian (\ref{heff}) is quadratic in the spin operators, and standard
spin wave  theory may be applied.\cite{mahan} The agreement of the
exact diagonalization data with the prediction of spin wave theory gives
further
support that the large spin limit is the natural tool to classify the
elementary
excitations  for this model.\cite{antimo}

In order to derive an effective hamiltonian which is consistent with
the previous results  I simplify the $H_p^{eff}$ in (\ref{heff})
by performing a systematic  expansion at long wavelength and large spin.
The large spin limit is in the same spirit of a low energy expansion.
 For instance the non-linear $\sigma$
model\cite{nelson}, which is widely accepted to represent the low energy
physics for the Heisenberg  antiferromagnet can be formally obtained as a
double
expansion in $1/s$ and at long wavelength.\cite{haldane,parola}

\section{ Large spin limit}
For large spin one can introduce Holstein-Primakov
 operators $a^{\dag}_R$ and
$b^{\dag}_R$ , describing a spin fluctuation over the N\'eel state in the
 $R \in A$
 or in the $R\in B$  sublattice of the antiferromagnet, respectively.
Due to the N\'eel AF order,
$a^{\dag}_R$ change the total spin $S_z$ by $+1$, conversely $b^{\dag}_R$
change it by $-1$. Here I assumed that the N\'eel state $|N>$ has
spin down in the $A$ sublattice and spin up in the $B$ one, the remaining
possibility being denoted by $|N^\prime>$, the ``anti''-N\'eel state, where
 of course $a^{\dag}_R$ and $b^{\dag}_R$ change the total spin $S_z$ in the
opposite way.

 The spin operator $\chi_{i,j}$ appearing in $H_{SW}$
, as well  as  in the general (\ref{heff}),
is extended  at  arbitrary spin with the
rotationally invariant expression \begin{equation}
\chi_{i,j} \to
\chi^{(s)}_{i,j} = {1 \over 2  s^2 } \vec S_i \,\cdot \,\vec S_j +  {1\over 2}
{}.
\label{chi}
\end{equation}
where the overall factors and constants
are set by the requirement that $\langle \chi^{(s)}_{i,j}\rangle$
is one or zero if the spins at sites $i$,$j$ are
parallel or antiparallel, respectively.
At large spin $\chi^{(\infty)}_{i,j} = {1\over 2 S}\left[ ( a^{\dag}_i + b_{j}
)
(a_i+  b^{\dag}_{j}) -1 \right] $ and, without the hole,   the
resulting large spin hamiltonian reads: $H_{SW}=J d/ (2 S)  \sum \limits_{k \in
BZ^{\prime}}  a^{\dag}_k a_k
+ b^{\dag}_k b_k + \gamma_k  (a^{\dag}_k b^{\dag}_{-k} + \,\,\,{\rm
h.c.})  $, where $\gamma_k=  {1 \over 2 d}  \sum \limits _{\tau_\mu} e^{i k
\tau_\mu} $ and the magnetic Brillouin zone ($BZ^\prime$)  is defined by the
 momenta $k$ satisfying  $\gamma_k \ge 0$.
 After a simple Bogoliubov transformation  one can express the
elementary excitations of the Heisenberg model at large spin in term of modes
with defined momentum in ($BZ^{\prime}$).
$\alpha^{\dag}_k=u_k a^{\dag}_k -v_k b_{-k}$ with $S_z=1$ and
$\beta^{\dag}_k=u_k b^{\dag}_k - v_k a_{-k}$ with $S_z=-1$ , where  $u_k=\sqrt{
1+\epsilon_k \over 2 \epsilon_k} $, $v_k= -\sqrt{1-\epsilon_k \over 2
\epsilon_k} $ and $\epsilon_k=\sqrt{1-\gamma_k^2}$ is proportional  to the
energy of  a spin wave excitation $E_k= J { d \over  (2 S)}   \epsilon_k$.

 The presence of the hole leads to further complications because the
large spin limit of the effective hamiltonian is not exactly solvable.
Anyway , at large spin , in the sector of  total spin $S_z=0$ and $S=0$,  it is
possible to show that the ground state of the hamiltonian (\ref{heff})  can be
written as a fluctuation operator $F$ acting over  the N\'eel state plus the
one
obtained  by replacing spin-up with spin down and vice versa ( a kind of
time-reversal symmetry\cite{siggia})  :  $$|\psi_H>= F(a^{\dag},b^{\dag})
|N>\pm
 F(a^{\dag},b^{\dag}) |N^{\prime}> + O({1\over S})$$, where the
negative or positive sign refer to $L/2$ odd  or even  respectively. For
simplicity we restrict in the following  to the latter case.
 For instance for the
Heisenberg ground state $|\psi_G>$, $F$ is gaussian at large spin and, apart
 for normalization, is given by:
$F \propto e^{\sum\limits_{k \in BZ^{\prime} }  {v_k \over u_k} a^{\dag}_k
b^{\dag}_{-k}} $.

The translation operator appearing in the effective
hamiltonian (\ref{heff}) is easily defined on any time-reversal symmetric state
$|\psi_H>$ as an operator acting  as
\begin{equation}
T_{\tau_\mu} a^{\dag}_k T_{-\tau_\mu} =e^{ i k \tau_\mu} b^{\dag}_k \,\,\, ,
\,\,\, T_{\tau_\mu} b^{\dag}_k T_{-\tau_\mu} =e^{ i k \tau_\mu} a^{\dag}_k
\label{trans} \end{equation}
 In other words any {\em small} fluctuation over the N\'eel state $|N>$ is
translated to a {\em small} fluctuation over $|N^{\prime}>$
 ($T_{\tau_\mu} |N> =|N^{\prime}>$ independent of $\tau_\mu$). The same
definition (\ref{trans}) holds by replacing  $a_k,b_k \to  \alpha_k, \beta_k$
because  $H_{SW}$ conserves the translation symmetry.
After this identification for the translation operator, using that the two
component of $|\psi_H>$ are orthogonal $o(e^{-S})$ for large spin, one can
 derive the  Scr\"oedinger equation satisfied by  $F|N>$  only, determined by
the hamiltonian $H^{eff}_p= \sum\limits_{\tau_\mu} \chi_{O,\tau_\mu}^{(\infty)}
(t e^{\vec p \cdot \vec \tau_\mu} T_{\tau_\mu} -J/2) + H_{SW}.$

\section{ Long-wavelength limit}
In this  limit the translation operator
defined by the previous relations (\ref{trans}) can be written as a product
 of a ``smooth'' operator $L_{\tau_\mu}$ and a discrete one $P$ defined by
 $P \alpha  P = \beta\,\,\, P \beta P =\alpha$, and such that
$P^{\dag}=P$ and $P^2=I$:
\begin{eqnarray}
T_{\tau_\mu}&=&L_{\tau_\mu} P \nonumber \\
L_{\tau_\mu} &=& e^{\sum\limits_{ k \in BZ^\prime } i \, \vec k \cdot
\tau_\mu  (
\alpha^{\dag}_k \alpha_k  + \beta^{\dag}_k \beta_k ) } \nonumber \\
P &=& e^{ i {\pi\over 2} \sum\limits_k (\alpha^{\dag}_k -\beta^{\dag}_k)
(\alpha_k -\beta_k)}  \nonumber
\end{eqnarray}
At long wavelength $L_{\tau_\mu}$ can be then expanded  for small
$k$\cite{aleandro}:
$L_{\tau_\mu}= I + \sum \limits_k i \, \vec k \cdot \tau_\mu  ( \alpha^{\dag}_k
\alpha_k  + \beta^{\dag}_k \beta_k )  + O(k^2) $.
A similar expansion is possible for the term $\chi_{i,j}$ appearing both
in $H_{SW}$ and in the local part of the hamiltonian. For $H_{SW}$ one then
obtains the correct low energy behavior  $H_{SW} = \sum\limits_k c |k|
(\alpha^{\dag}_k \alpha_k  + \beta^{\dag}_k \beta_k ) $, where $c= J
\sqrt{d}/(2 s)$ is the spin -wave  velocity
and the reference energy of the undoped system $H_{SW}$ is set to $0$ for
convenience.
Moreover the local term $\chi_{0,\tau_\mu}$  can be expanded  up to
$ O(k^2,{1\over s^2}) $ as:
$$\chi_{0,\tau_\mu} = { 1 \over 2 s} \left[ E^{\dag} E + O^{\dag}_{\mu} O_\mu +
i  ( O^{\dag}_{\mu} E - E^{\dag} O_\mu) -1 \right]$$
where:
$E^{\dag}  = \sqrt{1 \over \sqrt{d} \, L} \sum\limits_{k \in BZ^\prime}
\sqrt{|k|}  (\alpha_k+\beta_k) $ and

$O^{\dag}_{\mu} = \sqrt{\sqrt{d} \over L} \sum\limits_{k \in BZ^\prime} { \vec
k \cdot \vec \tau_{\mu} \over \sqrt{|k|} }  (\alpha_k^{\dag} + \beta_k)).$

After substituting
 the translation operator and $\chi_{i,j}$ in $H^{eff}_p$ (\ref{heff}) at
leading  order in $1/s$ and $k$ I consistently obtain  the following
hamiltonian which is quadratic in $\alpha_k$ and $\beta_k$, apart for  the
operator $P$, which cannot be simplified at long wavelength:
 \begin{eqnarray}
H_{t-J} &=& {J d  s (2s+1) \over 2 s^2} + H_{\rm LW} + H_{O} + O( k^2, 1/s^2)
\nonumber \\
 H_{\rm O}&=& {i \over 2 s} \sum\limits_{\tau_\mu}  (O^{\dag}_\mu E -
E^{\dag} O_{\tau_\mu}  )  t P e^{i p \tau_\mu} \nonumber \\
H_{\rm LW} &=& {-2 d t P \over 2s} \gamma_p  + {1 \over 2 s}
\sum\limits_{\tau_\mu} (E^{\dag} E + O_{\tau_\mu}^ {\dag} O_{\tau_\mu} ) ( t P
e^{i p \tau_\mu} -{J \over 2} ) + H^{\prime}_{\rm SW}
\label{lwt}
\end{eqnarray}
where $H^{\prime}_{SW}$ defines  a renormalization of the spin-wave velocity
due
to the  recoil of the hole in the AF background:
$$ H^{\prime}_{\rm SW} = {J d \over 2 s}  \sum\limits_{k\in BZ^\prime} \left[
{|k| \over \sqrt{d}}  +2 t P \vec  \nabla \gamma_p \cdot \vec k \right] (
\alpha_k^{\dag} \alpha_k + \beta_k^{\dag} \beta_k)
$$
Using that $P E^{\dag} P = E$ and $P O^{\dag}_{\mu} P = O_{\mu}$
it is simple  to show that $H_{LW}$ commutes with the operator $P$
 ($[E,E^{\dag}]=[O_{\mu},O^{\dag}_{\mu}]=0$), , while
  $H_O$ and the full hamiltonian  $H_{LW}+H_O$ do not.
 Indeed $H_{LW}+H_{O}$ determines all the eigenstates
at leading order $O(1)$ in $k$, and I will  show in the following that
$H_O$ gives  negligible  $O(k)$ corrections to the exact eigenstates
(or $O(k^2)$ to the energy) of $H_{LW}+H_O$, provided $\gamma_p \ne 0$.
In fact   $H_{LW}$
 has a gap  of order $t \gamma_p$ between
the two sectors differing by the  value of $P$, as indicated by the leading
$ \sim - t \gamma_p P$ $O(k^0)$-term in (\ref{lwt}).
 Then, by standard
perturbation theory, the effect of $H_O$ to the eigenstates of $H_{LW}$ is  of
higher order in $k$, just because of the mentioned gap and the fact that $H_O$
has matrix elements only between  opposite  sectors.

  At long wavelength
therefore one can  neglect  the odd term  $H_{O}$ and study
  $H_{LW}$ for $t\gamma_p P > 0$, which represents   the {\em exact} long
wavelength, large spin limit for one hole in the $t-J$ model.
 It is remarkable that  a   discrete operator,
$P$ which is not defined    at short distance becomes an
{\em  exact} symmetry  at long wavelength and is therefore important to
classify
the  elementary excitations of the model.
Roughly speaking this symmetry manifests itself
 at large distance from the hole where the two sublattices $A$ and $B$
are indistinguishable,  and we may  have even $P=1$ and odd $P=-1$ spin
fluctuations over the two sublattices.

\section{Consequences and Conclusions}
 The hamiltonian  $H_{LW}$ (\ref{lwt})
 commutes with $P$ and we can diagonalize it
in each sector where $P=\pm 1 $ is a definite quantum number. The
resulting hamiltonians $H_{LW,P=\pm 1}$,
 obtained by replacing $P$ with $\pm 1$ in $H_{LW}$
 are quadratic and of the standard Bogoliubov form, and admit both types of
eigenstates with   $P=1$ and $P=-1$, although their ground state is always
symmetric with $P=1$.
Thus at the end of the {\em exact}  analytical solution we have  to project
out the forbidden eigenstates with $P=\mp 1$.
The low energy hamiltonian $H_{LW}$ for $t \gamma_p P > 0$ is therefore exactly
solvable and it is found to be stable   for $J > \sim t$, i.e. without bose
condensation of modes with negative energy.
For small $J$ it is possible that the N\'eel semiclassical reference state has
to be replaced by a spiral or a polaronic solution.
 However, since all such type
 of solutions becomes N\'eel like at large distance, the low energy hamiltonian
$H_{LW}$,well defined for large $J$, should  be also representative even for
the small $J$ limit, at least for the  classification of the energy levels.

For $\gamma_p >0$
the ground state of the hamiltonian $H_{LW}$ has $P=1$ and  has the general
gaussian form:
 \begin{equation}
|\psi_H>= e^{\sum\limits_{k \in BZ^\prime}
B_{k,k^\prime} \alpha^{\dag}_k \beta^{\dag}_{k^\prime} } |0>_{\alpha,\beta}
\label{psigauss}
\end{equation}
where the symmetric  matrix $B_{k,k^\prime}$ depends on the momentum
$p$ of the hole and $|0>_{\alpha,\beta}$
is the vacuum of the spin waves, i.e. the ground state of $H_{SW}$.
$B_{k,k^\prime}$  is a dimensionless matrix $O(1)$in $k$.
Notice that, in order to determine the leading $s\to \infty$, $k \to 0$ ground
state  wavefunction , the hamiltonian has to be known up to first order in
$1/s$ and  in $k$ .

 The excitations  in the singlet subspace can be obtained  by applying
an even number of spin-wave
eigenmodes $\psi^{\dag}_{\sigma,j}$ $j=0,\cdots L/2-1$,  defined by the linear
eigenvalue equation:  $[ H_{LW, P=1},\psi^{\dag}_{\sigma,j}] =
\lambda_j \psi^{\dag}_{\sigma,j}$ where $\lambda_j\ge 0$ are the associated
energies.
The quasiparticles $\psi^{\dag}_{\sigma,j}$  carry spin
 $\sigma=\pm 1$ and are  of the
following form:
$\psi^{\dag}_{1,j}= \sum\limits_{k \in BZ^\prime} u_j (k) \alpha^{\dag}_k -
\sum v_j (k) \beta_k$,  $\psi^{\dag}_{\sigma,j}= P \psi^{\dag}_{-\sigma,j} P$.
 Due to $SU(2)$ symmetry there exist in general two  singular
 eigenmodes with  $\lambda_0 =0$, $\psi^{\dag}_{1,\lambda_0}=a^{\dag}_{k=0} +
b_{k=0}$ and $\psi^{\dag}_{-1,\lambda_0}=\psi_{1,\lambda_0}$, corresponding
 for large $s$ to the total spin raising or lowering
 operators, commuting with the hamiltonian.  A careful treatment of these
 singular  modes shows that they have to be simply dropped out in the
singlet subspace.\cite{zhong}
The coefficients $u_j(k)$ and $v_j (k) $ can be explicitly determined by simple
inspection of the eigenvalue equation, but their exact analytical form is not
 important in this context and will be published elsewhere.\cite{sorella}

For $\gamma_p<0$, i.e. outside the magnetic Brillouin zone, we have to work
in  the sector $P=-1$ and the low energy limit  is rather surprising.
 For instance   the ground state  {\em has not} a gaussian form like before
because the lowest state $|\psi>_{H}$ of $H_{LW,P=-1}$ has not
the correct  quantum number. Instead the allowed ground state
  is an antisymmetric excitation over the  forbidden  state  with $P=1$:
\begin{equation}
 |\psi_{H,P=-1}> = (\psi^{\dag}_{1,\lambda_1} \psi^{\dag}_{-1,\lambda_2}
- \psi^{\dag}_{1,\lambda_2} \psi^{\dag}_{-1,\lambda_1}) |\psi_H>
\label{psias}
\end{equation}
The antisymmetric combination is the only one
consistent with the $S_z=0$ (singlet) subspace and $P=-1$.
Since the energy of the lowest spin wave excitations
$\lambda_{1,2} \sim L^{-1/d}$, then, for $L\to \infty$,
the ground state  energy with momentum  $p$ outside $BZ^\prime$  coincides
 with the corresponding  one with momentum $p+Q$ inside
$BZ^\prime$, where $Q=(\pi,\pi,...)$ is the  magnetic wavevector.
The two states are however orthogonal in the long wavelength limit.



 In the singlet subspace  all possible eigenstates are a composition of spin
one
elementary  excitations  $\psi^{\dag}_{\pm 1,\lambda_j}$.
 The effect of these elementary excitations over the one hole state
 (\ref{psieff}) is somehow similar to
 dress the hole by a collective cloud of spin waves, recently
 proposed by Dagotto and Schrieffer.\cite{dagotto} In their approach however
they cannot distinguish the anomalous excitations (\ref{psias})
from the conventional Fermi liquid ones (\ref{psigauss}).

The quasiparticle weight $Z_p=|<\psi_G|\psi_H>|^2$  measures the pole in the
spectral  weight or in the angle resolved photoemission experiments. A finite
value  of $Z_p$ in the infinite volume limit  indicates that
the corresponding one  hole excitation is  Fermi liquid like.

The prediction of our long wavelength hamiltonian is that in a quantum
antiferromagnet we have a sort of Fermi surface -satisfying the Luttinger
theorem- where the $k$ values inside
the zone $BZ^\prime$ define conventional-spin $1/2$ and charge
one-  excitations, with $Z_p >0 $ and the ones  outside $BZ^\prime$
 which admit a composite ground state made up of two spin -one elementary
 excitations, and thus having its images  in the subspaces
with $S=0,1,2$.  $Z_p=0$ for this kind of excitations. I have verified these
predictions on an $18$-site lattice  where for $p$ around $(\pi,\pi)$
 and $t/J >0.6$ I  get $S=2$ as the true ground state  almost degenerate with
the $S=0$ one, which has an almost negligible quasiparticle weight (the $S=2$
one  is of course by symmetry orthogonal to $|\psi_H>$).

In conclusion this  new classification  suggests
that conventional Fermi liquid theory cannot be applied in a doped
antiferromagnet because the elementary excitations, modulo the operator $P$ are
in correspondence with the elementary excitations of the Heisenberg model.
At finite doping  the concept of  ``Heisenberg-Fermi liquid'' maybe  a more
appropriate description of the low energy physics.

\section*{ACKNOWLEDGMENTS}
I acknowledge useful correspondence of unpublished work
by  A. Parola who pointed out first the effective spin hamiltonian
(\ref{heff}) in 1991. I am   also grateful to  E.  Tosatti, M. Rice,
A. Angelucci, and D. Poilblanc  for useful discussions.
 This work is dedicated to the memory of my father Pompeo.

\end{document}